\documentclass[aps,superscriptaddress,showpacs,onecolumn,nofootinbib]{revtex4}
\usepackage{graphicx,epsfig}
\newcommand{\be}{\begin{equation}}
\newcommand{\ee}{\end{equation}}
\newcommand{\ba}{\begin{eqnarray}}
\newcommand{\ea}{\end{eqnarray}}
\newcommand{\di}{{\rm d}}
\newcommand{\la}{\langle}
\newcommand{\ra}{\rangle}
\begin{document}
\title{\boldmath 
	Collins effect and single spin azimuthal asymmetries\\
    	in the HERMES and COMPASS experiments}
\author{A.~V.~Efremov}\affiliation{
		Joint Institute for Nuclear Research, Dubna, 141980 Russia}
\author{K.~Goeke} 
\author{P.~Schweitzer}\affiliation{
		Institut f\"ur Theoretische Physik II, 
             	Ruhr-Universit\"at Bochum, D-44780 Bochum, Germany}
\date{September 2003}
\begin{abstract}
\noindent 
Predictions are made for single spin azimuthal asymmetries due to the Collins 
effect in pion production from semi-inclusive deeply inelastic scattering off 
transversely and longitudinally polarized targets for the HERMES and COMPASS 
experiments. 
The $x$-dependence of the asymmetries is evaluated using the parton 
distribution functions from the chiral quark-soliton model.
The overall normalization of the predicted asymmetries is determined by the 
information on the Collins fragmentation function extracted from previous 
HERMES data on azimuthal asymmetries $A_{UL}^{\sin\phi}$ from a 
{\sl longitudinally} polarized target. 
The single spin asymmetries $A_{UT}$ from the {\sl transversely} polarized 
proton target are found to be about $20\%$ for positive and neutral pions
both at HERMES and COMPASS.
For a {\sl longitudinally} polarized target we obtain for COMPASS
$A_{UL}^{\sin\phi}\sim 1\%$ and $A_{UL}^{\sin2\phi}\sim 3\%$.
\end{abstract}
\maketitle
\section{Introduction}
\label{Sect:Introduction}

Noticeable single spin azimuthal asymmetries\footnote{
    $U$ denotes the unpolarized beam. $L$ (below also $T$) denotes the
    longitudinal (and transverse) target polarization with respect to the
    beam. The superscript $\sin\phi$ characterizes the azimuthal
    distribution of the produced hadrons with respect to the direction
    of the exchanged virtual photon. The precise definitions are
    given in the Appendix.} 
$A_{UL}^{\sin\phi}$ have been observed by the HERMES collaboration in pion and
kaon electro-production in semi-inclusive deep-inelastic scattering (SIDIS) of
an unpolarized lepton beam off a longitudinally polarized proton or deuteron 
target \cite{Avakian:rr,Airapetian:1999tv,Airapetian:2001eg,Airapetian:2002mf}.
Recently the CLAS collaboration reported the measurement of the
azimuthal asymmetry $A_{LU}^{\sin\phi}$ from SIDIS of a polarized beam 
off an unpolarized target \cite{Avakian:2003pk}. Previously indications
for the azimuthal asymmetry $A_{UT}$ from SIDIS of an unpolarized
beam off a transversely polarized target were reported by the SMC
collaboration \cite{Bravar:rq}. 

Assuming factorization these single spin asymmetries can be
explained by the Collins \cite{Collins:1992kk} and Sivers effect
\cite{Sivers:1989cc} in terms of so far unexplored distribution
and fragmentation functions, namely the nucleon chirally odd
twist-2 transversity distribution  $h_1^a$ and twist-3
distribution functions $h_L^a$ and $e^a$ \cite{transversity}, the
Collins fragmentation function $H_1^{\perp a}$
\cite{Collins:1992kk,hand}, the chirally even Sivers distribution
function $f_{1T}^{\perp a}$
\cite{Sivers:1989cc,Collins:2002kn,Brodsky:2002cx,Belitsky:2002sm,Anselmino:1994tv}
(and/or transverse momentum weighted moments thereof
\cite{Mulders:1995dh,Boer:1997nt}). The $H_1^{\perp a}$ and
$f_{1T}^{\perp a}$ quantify the Collins and Sivers effect. The
former describes the left--right asymmetry in the fragmentation of
a transversely polarized quark into an unpolarized hadron; the
latter describes the distribution of unpolarized quarks in a
transversely polarized nucleon. Both are referred to as T-odd
since, if there were no interactions, these functions would be
forbidden by time reversal.

The HERMES data on single spin azimuthal asymmetries from SIDIS off a 
longitudinally polarized target
\cite{Avakian:rr,Airapetian:1999tv,Airapetian:2001eg,Airapetian:2002mf}
provide important indications that the mechanisms suggested by
Collins and Sivers \cite{Collins:1992kk,Sivers:1989cc} work, 
which makes them most exciting but also difficult to interpret. 
It is not clear which portion of the observed effect should be assigned 
to the Collins- and which to the Sivers mechanism. 
Moreover, numerous novel distribution- and
fragmentation functions complicate the analysis. Reasonable
descriptions of the HERMES data
\cite{Avakian:rr,Airapetian:1999tv,Airapetian:2001eg,Airapetian:2002mf}
using different assumptions and models were given in
Refs.~\cite{DeSanctis:2000fh,Anselmino:2000mb,Efremov:2000za,Ma:2002ns,Korotkov:1999jx,Efremov:2001cz,Efremov:2001ia}
in terms of the Collins effect only. Noteworthy, information on
the Sivers function gained from phenomenological description of
single spin asymmetries in $pp^\uparrow\to\pi X$ \cite{Anselmino:1994tv} 
indicates that neglecting the Sivers effect in the analysis of the HERMES 
experiment could be justified \cite{Efremov:2003tf}.

The understanding of the underlying phenomena is difficult also because so far
there is only one clear observable for target single spin asymmetries in SIDIS
with polarized targets, the $A_{UL}^{\sin\phi}$ asymmetry measured by HERMES
\cite{Avakian:rr,Airapetian:1999tv,Airapetian:2001eg,Airapetian:2002mf}.
Although at HERMES $A_{UL}^{\sin\phi}$ was measured in
electro-production of different hadrons from different targets --
providing valuable insights into the flavour dependence of the
process -- the observation of other independent observables which
allow to distinguish the Collins and Sivers effect is needed to
clarify the situation.

The azimuthal asymmetry $A_{UL}^{\sin2\phi}$ is such an
observable, for it is generated by the Collins effect only
\cite{Mulders:1995dh,Boer:1997nt}. Unfortunately, in the
kinematics of the HERMES experiment $A_{UL}^{\sin2\phi}$ was found
rather small and consistent with zero within (relatively large)
error bars
\cite{Avakian:rr,Airapetian:1999tv,Airapetian:2001eg,Airapetian:2002mf}.
This asymmetry could be accessed in the CLAS experiment, which
operates at somehow lower energies and higher luminosity than
HERMES. In the different kinematics of the CLAS experiment
$A_{UL}^{\sin2\phi}$ is expected to be larger than at HERMES and
measurable \cite{Efremov:2002ut},\footnote{
    The different kinematics and high luminosity at CLAS have already
    been explored to measure another asymmetry previously
    found consistent with zero at HERMES, namely the azimuthal asymmetry
    in SIDIS of a polarized beam off an unpolarized target,
    $A_{LU}^{\sin\phi}$. This asymmetry could be due to the Collins
    effect \cite{Mulders:1995dh,Boer:1997nt} and provide first indications
    to the twist-3 distribution function $e^a(x)$ \cite{Efremov:2002ut}.}
and, indeed, encouraging preliminary results have already been reported in 
Ref.~\cite{Avakian-talk}. Also in the COMPASS experiment $A_{UL}^{\sin2\phi}$ 
will probably be observable -- as we will estimate below.

More conclusive insights, however, are expected from SIDIS
experiments with {\sl transversely} polarized targets\footnote{A
first observation of single spin azimuthal asymmetries in SIDIS
from a transversely polarized target -- which unfortunately
retained its preliminary status -- was reported from the SMC
experiment \cite{Bravar:rq}.}, where the Collins and Sivers
effects \cite{Collins:1992kk,Sivers:1989cc}, can cleanly be
distinguished \cite{Boer:1997nt}.  Those experiments are presently 
in progress at HERMES \cite{Makins:uq} and COMPASS \cite{LeGoff:qn}.
Estimates of these asymmetries for HERMES were presented
in Refs.~\cite{Korotkov:1999jx,Bacchetta:2002tk}.

In this paper we will predict the azimuthal single spin asymmetry
due to the Collins effect from a transversely polarized target for
the kinematics of the HERMES and COMPASS experiments. For that we
shall use predictions for the transversity distribution function
$h_1^a(x)$ from the chiral quark-soliton model \cite{h1-model} and
information on the analyzing power $\la H_1^\perp\ra/\la D_1\ra$ 
from a previous analysis~\cite{Efremov:2001cz} of the HERMES
data.\footnote{
	Actually, in that analysis ~\cite{Efremov:2001cz} the Sivers function 
	was neglected, which has later been shown to be theoretically 
	consistent and phenomenologically justified \cite{Efremov:2003tf}.} 
Indeed, the present approach,
based on the chiral-quark soliton model and the instanton vacuum picture, 
describes in a theoretically consistent and phenomenologically satisfying 
way \cite{Efremov:2001cz,Efremov:2001ia} the $x$-dependence of the HERMES 
data \cite{Avakian:rr,Airapetian:1999tv,Airapetian:2001eg,Airapetian:2002mf}.
In a certain sense the analyzing power $\la H_1^\perp\ra/\la D_1\ra$ 
from \cite{Efremov:2001cz} quantifies the amount of Collins effect 
needed to understand the HERMES data
\cite{Avakian:rr,Airapetian:1999tv,Airapetian:2001eg,Airapetian:2002mf}
within this approach. Therefore the comparison of our prediction
to the outcome of the HERMES and COMPASS transverse target
polarization experiments will yield more than an important test of
the approach and its consistency. An agreement would support also
the conclusion of Ref.~\cite{Efremov:2003tf} that the Sivers
effect can be neglected in $A_{UL}^{\sin\phi}$-asymmetries and it
would justify, a posteriori, the attempts \cite{DeSanctis:2000fh,Anselmino:2000mb,Efremov:2000za,Ma:2002ns,Korotkov:1999jx,Efremov:2001cz,Efremov:2001ia}
to understand HERMES data on $A_{UL}$ in terms of the Collins effect only.

The paper is organized as follows. In
Section~\ref{Sect:Collins-AUT} the SIDIS process and its
description is discussed under the assumption of factorization. In
Section~\ref{Sect:Ingredients} our assumptions on the novel
distribution and fragmentation functions are described. In
Sections~\ref{Sect:HERMES-AUT} and \ref{Sect:COMPASS-AUT-AUL} the
predictions are presented for the HERMES and COMPASS transverse
target polarization experiments, as well as for the longitudinal
target polarization experiment at COMPASS. In
Section~\ref{Sect:Sivers-effect} we present general comments on
the Sivers effect in SIDIS asymmetries.
Section~\ref{Sect:Conclusions} contains the summary and
conclusions.

\section{The contribution of the Collins effect to the azimuthal 
	asymmetry from a transversely polarized target}
\label{Sect:Collins-AUT}

In the HERMES and COMPASS experiments the cross sections
$\sigma_N^{\uparrow\downarrow}$ for the process
$lN^{\uparrow\downarrow}\rightarrow l'h X$ will be measured, where
$N^{\uparrow\downarrow}$ denotes the transversely with respect to
the beam polarized target, see Fig.~1. 

%
%
\begin{figure}
        \includegraphics[height=.16\textheight]{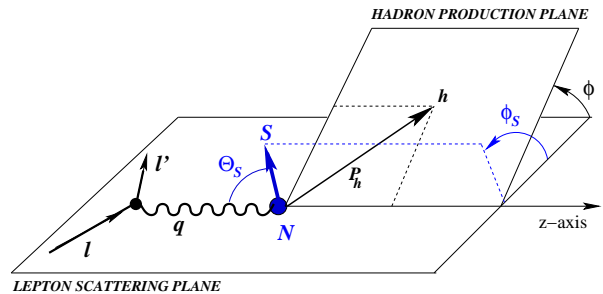}
        \caption{\footnotesize\sl
    	Kinematics of the process $lN^\uparrow\rightarrow l'h X$
    	in the lab frame.}
\end{figure}
%
%

With $P$, $l$ and $l'$ denoting the momenta of the target, incoming and
outgoing lepton the kinematic variables are defined as $s:=(P+l)^2$,
$q:= l-l'$ with $Q^2:= - q^2$,  and $W^2:= (P+q)^2$, and
\be\label{notation-1}
        x := \frac{Q^2}{2Pq}    \;,\;\;\;
        y := \frac{Pq}{Pl}      \;,\;\;\;
        z := \frac{PP_h}{Pq\;}  \;.\ee
Let $S^{\uparrow\downarrow}$ denote the modulus of the polarization 
vector. The component of the target polarization vector which is
{\sl transverse with respect to the hard photon} is characterized
by the angle $\Theta_S$, see Fig.~1, given by
	\be\label{spin-projection}
    	\sin\Theta_S
	=\frac{S^{\uparrow\downarrow}_T}{S^{\uparrow\downarrow}}
    	=\cos\theta_\gamma\sqrt{1+{\rm tan}^2\theta_\gamma\sin^2\phi_{S'}}
	\approx \cos\theta_\gamma \;,\ee
	where $\phi^\prime_S$ is the azimuthal angle of the target spin 
	direction around the lepton beam direction counted from the 
	scattering plane, and $\cos\theta_\gamma$ is given by
	\be\label{spin-projection-a}
	\cos\theta_\gamma=
	\sqrt{1-\frac{(4{M_N}^2x^2)(1-y-{M_N}^2x^2y^2/Q^2)}{(Q^2+4{M_N}^2x^2)}} \;.
	\ee	
	Since ${\rm tan}^2\theta_\gamma={\cal O}(M^2/Q^2)$ the approximation
 	in the last step of Eq.~(\ref{spin-projection-a}) works well.

With $\phi$ ($\phi_S$) denoting the azimuthal angles between the hadron
production plane (the nucleon spin) and the lepton scattering plane,
see Fig.~1, the observables of interest are defined as
\ba
        A_{UT}^{\sin(\phi+\phi_S)}(x) &=&
        \frac{  \displaystyle \int\!\!\di z\,\di y\,\di^2{\bf P}_{h\perp}\,
        \sin(\phi+\phi_s)\,\left(\frac{1}{S^\uparrow}\,
        \frac{\di^5\sigma^\uparrow}{\di x\,\di y\,\di z\,
        \di^2{\bf P}_{h\perp}} - \frac{1}{S^\downarrow}\,
        \frac{\di^5\sigma^\downarrow}{\di x\,\di y\,\di z\,
        \di^2{\bf P}_{h\perp}}\right)}
    {\;\;\;\;\;\;\;\;\;\;\;\displaystyle
        \frac{1}{2}\int\!\!\di z\,\di y\,\di^2{\bf P}_{h\perp}\,\left(
   \frac{\di^5\sigma^\uparrow  }{\di x\,\di y\,\di z\,\di^2{\bf P}_{h\perp}}+
   \frac{\di^5\sigma^\downarrow}{\di x\,\di y\,\di z\,\di^2{\bf P}_{h\perp}}
    \right)} \;,\label{AUT-without-kT}  \\
    && \phantom{X}\nonumber\\
        A_{UT}^{\sin(\phi+\phi_S)k_\perp/\la P_{h\perp}\ra}(x) &=&
        \frac{  \displaystyle \int\!\!\di z\,\di y\,\di^2{\bf P}_{h\perp}\,
        \sin(\phi+\phi_s)\,\frac{k_\perp}{\la P_{h\perp}\ra}\,
        \left(\frac{1}{S^\uparrow}
        \,\frac{\di^5\sigma^\uparrow}{\di x\,\di y\,\di z\,
        \di^2{\bf P}_{h\perp}} - \frac{1}{S^\downarrow}\,
        \frac{\di^5\sigma^\downarrow}{\di x\,\di y\,\di z\,
        \di^2{\bf P}_{h\perp}}\right)}
           {\;\;\;\;\;\;\;\;\;\;\;\;\;\;\;\;\;\displaystyle
           \frac{1}{2}\int\!\!\di z\,\di y\,\,\di^2{\bf P}_{h\perp}\,\left(
   \frac{\di^5\sigma^\uparrow  }{\di x\,\di y\,\di z\,\di^2{\bf P}_{h\perp}}+
   \frac{\di^5\sigma^\downarrow}{\di x\,\di y\,\di z\,\di^2{\bf P}_{h\perp}}
    \right)} \;.\nonumber\\
    && \label{AUT-with-kT}\ea
The weight $\sin(\phi+\phi_s)$ in Eq.~(\ref{AUT-without-kT}) has the
drawback to leave convoluted the transverse momenta in the unintegrated
distribution and fragmentation functions -- in this case 
$h_1(x,P_{{\rm N}\perp}^2)$ and $H_1^\perp(z,k_T^2)$ \cite{Mulders:1995dh}.
(For the meaning and definition of unintegrated distribution functions
in QCD see \cite{Collins:2003fm} and references therein.)
The additional power of transverse momentum\footnote{
    \label{footnote:notation}
    We use the notation of \cite{Mulders:1995dh,Boer:1997nt} with
    $H_1^\perp$ normalized to $\la P_{h\perp}\ra$ instead of $m_h$.
    Correspondingly we choose $\la P_{h\perp}\ra$ to compensate the
    dimension of $k_\perp$ in Eq.~(\ref{AUT-with-kT}).}
$k_\perp = |{\bf P}_{h\perp}|/z$ in the weight in
Eq.~(\ref{AUT-with-kT}) yields expressions where the transverse
momenta are disentangled in a model independent way
\cite{Boer:1997nt}.

Though the asymmetry weighted with $k_\perp$ in
Eq.~(\ref{AUT-with-kT}) is preferable from a theoretical point of
view \cite{Boer:1997nt}, we shall consider both asymmetries,
Eq.~(\ref{AUT-without-kT}) and Eq.~(\ref{AUT-with-kT}).
Considering also the  asymmetry (\ref{AUT-without-kT}) will allow
us to directly compare the predicted effect to the
$A_{UL}^{\sin\phi}$ asymmetries measured at HERMES
\cite{Avakian:rr,Airapetian:1999tv,Airapetian:2001eg,Airapetian:2002mf}
which were analyzed in a way analogous to
Eq.~(\ref{AUT-without-kT}).

The expressions for the differential cross sections entering the
asymmetries in Eqs.~(\ref{AUT-without-kT},~\ref{AUT-with-kT}) were
derived in \cite{Mulders:1995dh} assuming factorization. In order
to deconvolve the transverse momenta in
$A_{UT}^{\sin(\phi+\phi_S)}$ in Eq.~(\ref{AUT-without-kT}) we
assume the distributions of transverse momenta in the unintegrated
distribution and fragmentation functions to be Gaussian. This
ansatz is in fair agreement with the HERMES data in the case of
$A_{UL}^{\sin\phi}$ asymmetries
\cite{Avakian:rr,Airapetian:1999tv,Airapetian:2001eg,Airapetian:2002mf}.
Under this assumption one obtains \cite{Mulders:1995dh} (cf.\ also
\cite{Efremov:2001cz}) \be\label{AUT-without-kT-fin}
        A_{UT}^{\sin(\phi+\phi_s)}(x) = a_{\rm Gauss}\,B_T(x)\;
        \frac{\sum_a e_a^2\,x\, h_1^a(x)\,\la H_1^{\perp a}\ra}
             {\sum_b e_b^2\,x\, f_1^b(x)\,\la D_1^b\ra\,} \;,\ee
while the result for the $k_\perp$-weighted asymmetry is given by
\cite{Boer:1997nt}
\be\label{AUT-with-kT-fin}
        A_{UT}^{\sin(\phi+\phi_s)k_\perp/\la P_{h\perp}\ra}(x) = B_T(x)\;
        \frac{\sum_a e_a^2\,x\, h_1^a(x)\,\la H_1^{\perp (1)a}\ra}
             {\sum_b e_b^2\,x\, f_1^b(x)\,\la D_1^b\ra\,}\;, \ee
where $B_T(x)$ and $a_{\rm Gauss}$ are defined as (experimental cuts have
to be considered in the integration over $y$)
\ba\label{prefactor-B_T}
    B_T(x) &=& \frac{2\int\!\di y\,(1-y)\,\sin\Theta_S/Q^4}
    {\int\!\di y\,(1-y+y^2/2)\,/Q^4}\;, \\
   \label{a_Gauss}
    a_{\rm Gauss} &=& \frac{1}{2\la z\ra\sqrt{1+
    \la z^2\ra\la P_{{\rm N}\perp}^2\ra/\la P_{h\perp}^2\ra}} \;, \ea
where $\la P_{{\rm N}\perp}^2\ra$ and $\la P_{h\perp}^2\ra/\la z^2\ra$ are the
mean transverse momentum squares characterizing the Gaussian distributions of
transverse momenta in the unintegrated distribution and fragmentation function.
The prefactor $a_{\rm Gauss}$ contains the model dependence; it would be
different if we assumed the distributions of transverse momenta to be
different from Gaussian.
$H_1^{\perp (1)a}(z)$ in Eq.~(\ref{AUT-with-kT-fin}) is defined by
\cite{Boer:1997nt} (cf.\ footnote~\ref{footnote:notation})
\be\label{weighted-H1perp}
    H_1^{\perp (1) a}(z)= \int\di^2{\bf k}_T\;
    \frac{z^2{\bf k}_T^2}{2\la P_{h\perp}\ra^2}\;
    H_1^{\perp a}(z,z^2{\bf k}_T^2)\;.\ee

\section{Transversity distribution and Collins fragmentation function}
\label{Sect:Ingredients}

In order to estimate 
the azimuthal asymmetries, Eqs.~(\ref{AUT-without-kT},~\ref{AUT-with-kT}) and
(\ref{AUT-without-kT-fin},~\ref{AUT-with-kT-fin}), one has to know $h_1^a$ and 
$H_1^{\perp a}$. For the former we shall use the predictions of the chiral 
quark-soliton model ($\chi$QSM) \cite{h1-model}, and for the latter our 
analysis of the HERMES data from Ref.~\cite{Efremov:2001cz}.

\paragraph*{Chirally and T-odd distribution functions.}
The $\chi$QSM is an effective relativistic quantum
field-theoretical model with explicit quark degrees of freedom, in
which twist-2 nucleon distribution functions can unambiguously be
defined and evaluated at a low renormalization point of about
$(600-700)$ MeV. The $\chi$QSM has been derived from the instanton
model of the QCD vacuum \cite{Diakonov:2002fq} and has been shown
to describe well numerous static nucleonic observables without
adjustable parameters \cite{Christov:1995vm}. The
field-theoretical nature of the model is crucial to ensure the
theoretical consistency of the approach: the quark and antiquark
distribution functions computed in the model satisfy all general
QCD requirements \cite{Diakonov:1996sr}. The results of the model
agree for the distribution functions $f_1^a(x)$, $g_1^a(x)$ and
$g_T^a(x)$ \cite{Diakonov:1996sr,Diakonov:1997vc,Balla:1997hf}
within (10 - 30)$\%$ with phenomenological information. This
encourages confidence that the model describes the nucleon
transversity distribution function $h_1^a(x)$ \cite{h1-model} with
a similar accuracy.

In the following we will need also the deuteron transversity
distribution. Corrections due to the D-state admixture were
estimated to be very similar to the corresponding corrections in
the case of the helicity distribution function
\cite{Umnikov:1996hy}. Since these corrections are smaller than
other theoretical uncertainties in our approach we shall disregard
them here and estimate e.g. for the $u$-quark \be\label{h1-deut}
    h_1^{u/D}(x) \approx h_1^{u/p}(x) + h_1^{u/n}(x)
    = h_1^u(x) + h_1^d(x) \;,
\ee
where isospin symmetry was used in the last step, and $h_1^u(x)$ and
$h_1^d(x)$ refer (as always) to the proton.

In the $\chi$QSM $h_1^a(x)\neq g_1^a(x)$ already at a low
normalization point. However, due to the large error bars the
present data do not discriminate between different models.
Therefore reasonable descriptions of the $A_{UL}^{\sin\phi}$
asymmetries have also been obtained assuming $h_1^a(x) = g_1^a(x)$
being motivated by the non-relativistic quark model or using other
models \cite{Ma:2002ns}. An advantage of relying on predictions
from the $\chi$QSM (based on the instanton vacuum) lies in the
fact that all novel distribution functions are taken from an
approach, which is internally consistent and which has been shown
in many different observables to be reasonable. For example, in
the instanton vacuum model the pure twist-3 contribution
$\widetilde{h}_L^a(x)$ to $h_L^a(x)$ is strongly suppressed
\cite{Dressler:1999hc}. Thus in this approach one can justifiably
approximate $h_L^a(x)$ by its twist-2 (``Wandzura-Wilczek'' like)
term $h_L^a(x)= 2x \int_x^1 \di
x^\prime\,h_1^a(x^\prime)/x^{\prime 2}$. Moreover, T-odd
distribution functions vanish in the $\chi$QSM (as they do in a
large class of other chiral soliton models)
\cite{Pobylitsa:2002fr}. Therefore in this approach it is consistent 
to neglect the Sivers effect in $A_{UL}^{\sin\phi}$ asymmetries 
\cite{Efremov:2003tf}, see also Section~\ref{Sect:Sivers-effect} 
for further comments.

In Refs.~\cite{Efremov:2001cz,Efremov:2001ia} the present approach
has been shown to describe well the $x$-dependence of $A_{UL}$
from the HERMES longitudinal target polarization experiments
\cite{Avakian:rr,Airapetian:1999tv,Airapetian:2001eg,Airapetian:2002mf}.

\paragraph*{The Collins fragmentation function.}
Let us define the favoured Collins fragmentation function as
\be\label{Def:H1perp-favoured}
	H_1^{\perp} \equiv 
	H_1^{\perp u/\pi^+} =  H_1^{\perp\bar d/\pi^+} = 
	H_1^{\perp d/\pi^-} = 2H_1^{\perp u/\pi^0}     = \dots\;\mbox{etc.}\gg
	H_1^{\perp d/\pi^+} = H_1^{\perp\bar u/\pi^+}  = \dots\;\mbox{etc.}
\ee
The equalities in Eq.~(\ref{Def:H1perp-favoured}) follow from charge
conjugation and isospin symmetry. The strong suppression of the unfavoured
with respect to the favoured Collins fragmentation function has been concluded
on  the basis of the  Sch\"afer-Teryaev sum rule \cite{Schafer:1999kn}. 

In Ref.~\cite{Efremov:2001cz} information on $H_1^\perp$ was
gained from the HERMES data on the $A_{UL}^{\sin\phi}$ asymmetry
in $\pi^+$ and $\pi^0$ production
\cite{Airapetian:1999tv,Airapetian:2001eg}. For that the
transverse momentum distributions were assumed to be Gaussian and
the parton distribution functions $h_1^a$ and $h_L^a$ were taken
from the chiral quark soliton model. For the analyzing power the
value was found ($D_1$ denotes the favoured unpolarized
fragmentation function) \be\label{apower}
    \la H_1^\perp\ra/\la D_1\ra=(13.8\pm 2.8)\%
\ee
at $\la z\ra = 0.4$ and $\la Q^2\ra = 2.5\;{\rm GeV}^2$ \cite{Efremov:2001cz}.
The result in Eq.~(\ref{apower}) contains -- apart from the shown
statistical error from the HERMES experiment -- further uncertainties due to
the systematic error of the HERMES experiment and model dependence.
These errors need not be considered in the following, when the above
result is used to make predictions for $A_{UT}^{\sin(\phi+\phi_S)}$ in the
HERMES experiment in combination with results from the chiral quark-soliton
model and the instanton vacuum model.
In a certain sense the result in Eq.~(\ref{apower}) can be considered as a fit
to the $A_{UL}^{\sin\phi}$ data \cite{Airapetian:1999tv,Airapetian:2001eg}.
Noteworthy, a result numerically close to Eq.~(\ref{apower}) was obtained 
in the model calculation of Ref.~\cite{Bacchetta:2002tk}. 

In $e^+e^-$ annihilation the Collins effect can give rise to a specific
azimuthal asymmetry of a hadron in a jet around the axis in the direction
of the second hadron in the opposite jet.
This asymmetry was measured using the DELPHI data collection and a value
$|{\la H_1^\perp\ra/\la D_1\ra}|=(12.5\pm 1.4)\%$ for $\la z\ra\simeq 0.4$
at a scale of $M_Z^2$ was reported \cite{todd,czjp99}.\footnote{
    This result is referred to as ``more optimistic'' since it is subject
    to presumably larger systematic uncertainties than the ``more
    reliable'' value $|{\la H_1^\perp\ra/\la D_1\ra}|=(6.3\pm 2.0)\%$
    reported in \cite{todd,czjp99} which has presumably smaller systematic
    errors. For both values no estimate of systematic errors could
    be given in \cite{todd,czjp99}.}
In previous works \cite{Efremov:2001cz,Efremov:2001ia} this value
(assuming a positive sign) was used to analyze the HERMES data
\cite{Avakian:rr,Airapetian:1999tv,Airapetian:2001eg,Airapetian:2002mf}.
For that the scale dependence of the ratio ${\la H_1^\perp\ra/\la D_1\ra}$ was
assumed to be weak and possible Sudakov suppression effects \cite{Boer:2001he}
were neglected. However, as shown in Ref.~\cite{Boer:2003cm} the Collins
fragmentation function could be process-dependent, i.e.\  different in
$e^+e^-$ annihilation and SIDIS.

Therefore, in this note we shall use the result in Eq.~(\ref{apower})
extracted from SIDIS HERMES data. Numerically the difference is not relevant
-- from a theoretical point of view, however, the use of the result in
Eq.~(\ref{apower}) is preferable for our purpose to describe SIDIS processes.

\section{\boldmath Collins $A_{UT}$ asymmetries in the HERMES experiment}
\label{Sect:HERMES-AUT}

\paragraph*{The asymmetry $A_{UT}^{\sin(\phi+\phi_S)}$.}
In order to estimate $A_{UT}^{\sin(\phi+\phi_S)}$ in the HERMES
experiment we rely on the same assumptions and approximations
which were used in Refs.~\cite{Efremov:2001cz,Efremov:2001ia} to
analyze the HERMES data on the $A_{UL}^{\sin\phi}$ asymmetries. In
particular we assume a Gaussian distribution of transverse momenta
(cf.\  Section~\ref{Sect:Collins-AUT}), take $h_1^a(x)$ from the
$\chi$QSM and $H_1^\perp$ from our previous analysis of
HERMES-data \cite{Efremov:2003tf} and assume favoured
fragmentation -- as described in Section~\ref{Sect:Ingredients}. For
the unpolarized distribution functions $f_1^a(x)$ we use the
parameterization of Ref~\cite{GRV}. For the
parameter characterizing the (Gaussian) distribution of transverse momenta in 
the nucleon we shall use the estimate $\la P_{{\rm N}\perp}\ra=0.4\,{\rm GeV}$
from Refs.~\cite{Martin:1998sq,Abreu:1996na}. The result, however, is only 
weakly sensitive to the actual choice for this parameter.

The beam in the HERMES experiment has an energy of
$E_{\rm beam}=26.7\,{\rm GeV}$.
We assume the cuts implicit in the integrations in Eq.~(\ref{prefactor-B_T})
to be the same as in the longitudinal target polarization experiments
\cite{Avakian:rr,Airapetian:1999tv,Airapetian:2001eg,Airapetian:2002mf}
\be\label{exp-cuts-HERMES}
        1\,{\rm GeV}^2 < Q^2 < 15\,{\rm GeV}^2  , \;\;\;
        2\,{\rm GeV} < W                        , \;\;\;
        0.2 < y < 0.85                        \;, \;\;\;
        0.023 < x < 0.4                       \;,\ee
and $0.2<z<0.7$ with $\la z\ra=0.4$, and $\la
P_{h\perp}\ra=0.4\;{\rm GeV}$. Note that strictly speaking we
neglect the implicit dependence of distribution and fragmentation
functions on $y$ through the scale $Q^2=xy(s-{M_N}^2)$, and evaluate
them instead at the average scale in the HERMES experiment $\la
Q^2\ra=2.5\,{\rm GeV}^2$. The predictions for
$A_{UT}^{\sin(\phi+\phi_S)}$ for the transversely polarized
proton and deuterium target are shown in Figs.~2a and 2b,
respectively.

Figs.~2a and 2b demonstrate that $A_{UT}^{\sin(\phi+\phi_S)}$ is
sizeable, roughly $20\%$ for positive and neutral pions for the
proton target and about $10\%$ for all pions for the deuteron target. 
Comparing this result with the $A_{UL}^{\sin\phi}$ asymmetries $\sim(2-4)\%$
\cite{Avakian:rr,Airapetian:1999tv,Airapetian:2001eg,Airapetian:2002mf}
we see that $A_{UT}^{\sin(\phi+\phi_S)}$ asymmetry can clearly be observed, 
cf.\  \cite{Korotkov:1999jx}. A comparably large value for this asymmetry was
estimated in \cite{Bacchetta:2002tk} on the basis of a model calculation for 
the Collins function and assuming $h_1^a(x)$ to saturate the Soffer bound
\cite{Soffer:1995ww}.

The accuracy of the predictions --
for $\pi^+$ and $\pi^0$ asymmetries from a proton target -- is
mainly determined by the theoretical uncertainty of the $\chi$QSM
prediction for $h_1^a(x)$ of about $20\%$ and the statistical
error of the analyzing power (\ref{apower}) from the HERMES
experiment.
For negative pions from a proton, however, there might be additional sizeable
corrections due to unfavoured flavour fragmentation \cite{Ma:2002ns}.
In this case the small unfavoured Collins fragmentation function is multiplied
by the large $\frac49 h_1^u(x)$ while the large favoured fragmentation function
is multiplied by the small $\frac19 h_1^d(x)$.\footnote{
    The antiquark distributions can be disregarded for this
    qualitative consideration. The same applies to unpolarized
    fragmentation. Since $f_1^a(x)$ and $D_1^a(z)$ are positive,
    the effect of unpolarized unfavoured fragmentation may decrease
    the asymmetry but cannot change its sign -- as could do the
    polarized unfavoured fragmentation in the case of $\pi^-$
    from a proton target \cite{Ma:2002ns}.}
Therefore $\pi^-$ is more sensitive to corrections due to unfavoured
fragmentation than $\pi^+$ and $\pi^0$ where $u$-quark dominance
($h_1^u(x)\gg |h_1^d(x)|$) tends to enhance the favoured fragmentation effect.
Similar reservations apply to the deuteron target where there is no $u$-quark 
dominance -- apart from that introduced by the quark electric charges.

In Ref.~\cite{Efremov:2001ia} $A_{UL}^{\sin\phi}$ asymmetries for kaons
have been estimated assuming that the analyzing power for kaons
is approximately equal to that of pions, i.e.\
$\la H_1^\perp\ra/\la D_1\ra|_{\rm kaon} \approx $
$\la H_1^\perp\ra/\la D_1\ra|_{\rm pion}$. This relation would hold exactly in
the chiral limit (where pions and kaons would be massless Goldstone bosons).
The kaon $A_{UL}^{\sin\phi}$ asymmetries predicted in \cite{Efremov:2001ia}
on the basis of this assumption compare well with the HERMES data within the
(admittedly rather large) statistical error \cite{Airapetian:2002mf}.
Under this assumption one could expect for the transverse target polarization
experiment (cf.\  Ref.~\cite{Efremov:2001ia} for further details)
\ba
&&  A_{UT}^{\sin(\phi+\phi_S)}(K^+) \approx
    A_{UT}^{\sin(\phi+\phi_S)}(K^0) \approx
    A_{UT}^{\sin(\phi+\phi_S)}(\pi^+)\;,\nonumber\\
&&
    A_{UT}^{\sin(\phi+\phi_S)}(\bar{K}^0) \approx
    A_{UT}^{\sin(\phi+\phi_S)}(K^-) \approx 0 \; .\label{AUT-kaons} \ea

%
%
\begin{figure}[t!]\label{Fig:predictions-AUT-HERMES}
\begin{tabular}{cc}
    \includegraphics[width=7.5cm,height=7.5cm]{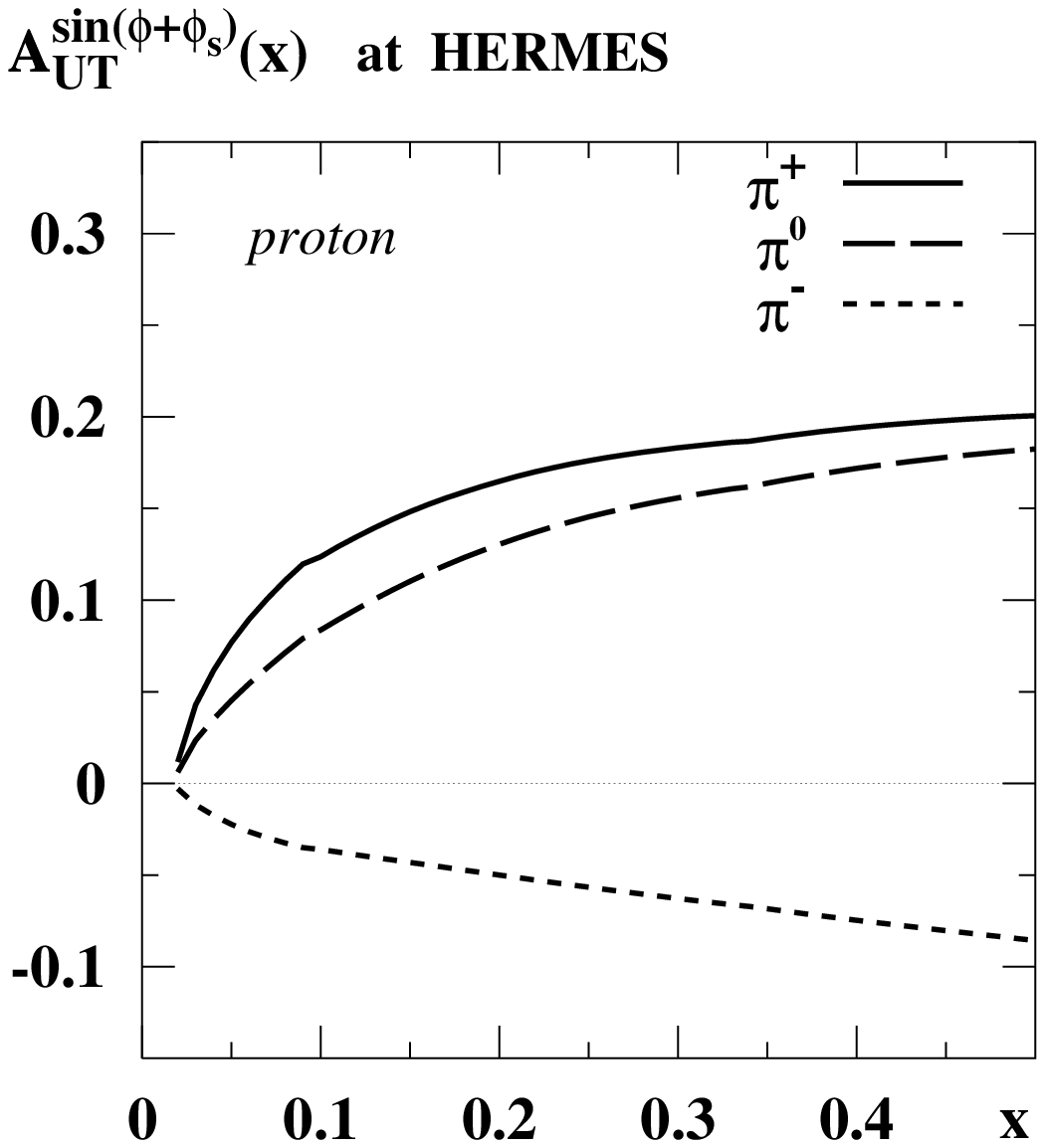}&
    \includegraphics[width=7.5cm,height=7.5cm]{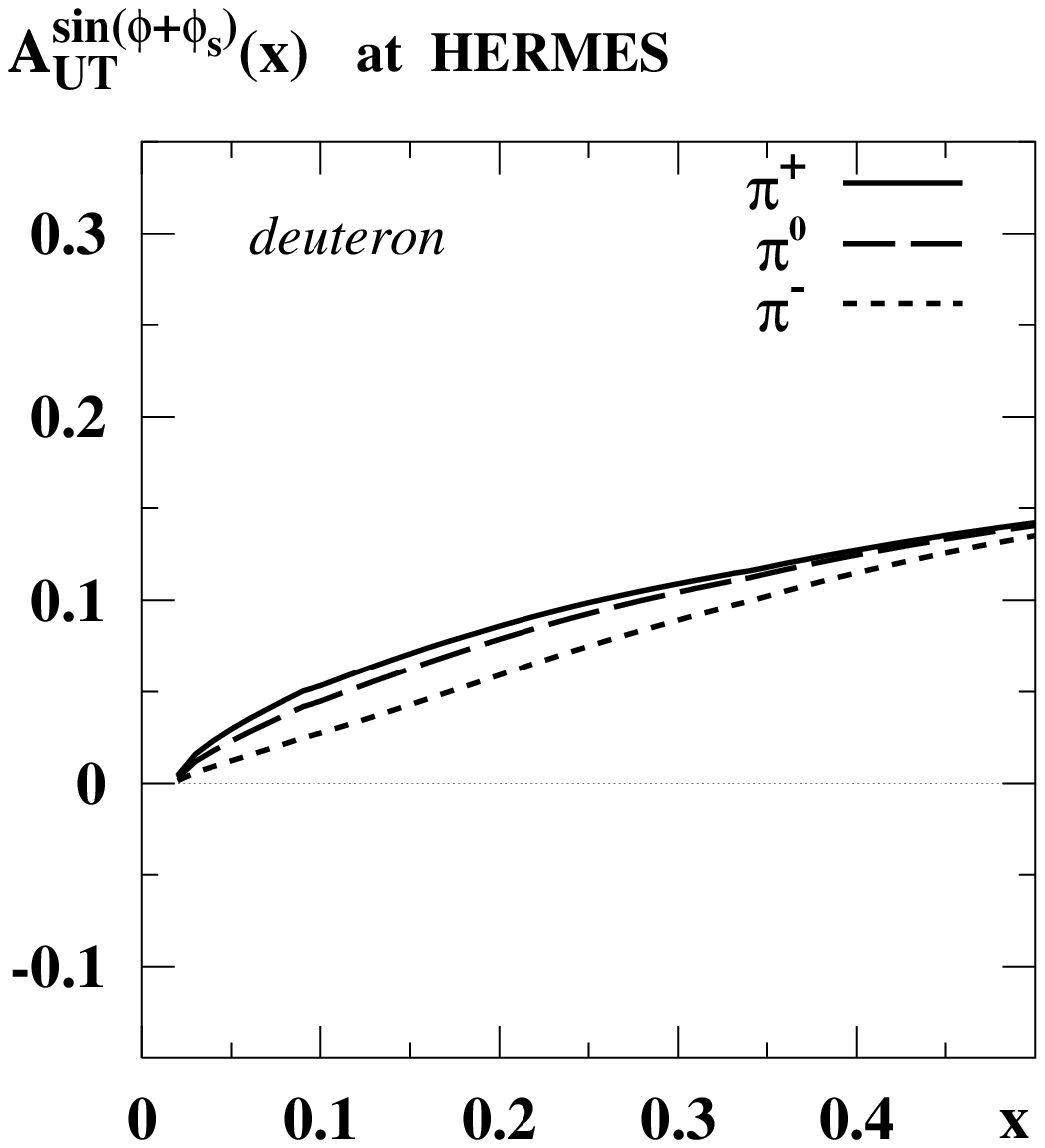}\cr
    \hspace{0.8cm} {\bf a} & \hspace{0.8cm} {\bf b}
\end{tabular}
    \caption{\footnotesize\sl
    Predictions for azimuthal asymmetries $A_{UT}^{\sin(\phi+\phi_S)}(x)$
    in SIDIS pion productions from transversely polarized proton (a) and
    deuteron (b) targets for kinematics of the HERMES experiment.} 
\end{figure}
%
%

\paragraph*{The asymmetry
       $A_{UT}^{\sin(\phi+\phi_S)k_\perp/\la P_{h\perp}\ra}$.}
In this case we need the transverse momentum weighted moment of
the favoured Collins fragmentation function $H_1^{\perp (1)}$,
see Eq.~(\ref{AUT-with-kT-fin}).
Under the assumption of a Gaussian distribution of transverse momenta in
Eq.~(\ref{weighted-H1perp}) one obtains for $H_1^{\perp (1)}$ averaged
over $z$
\be\label{apower-weighted}
    \la H_1^{\perp (1)}\ra
    =\frac{\la P_{h\perp}^2\ra}{2\la P_{h\perp}\ra^2}\;
    \la H_1^\perp\ra
    =\frac{2}{\pi} \;\la H_1^\perp\ra \;,\ee
where we used the relation $\la P_{h\perp}^2\ra/\la P_{h\perp}\ra^2=4/\pi$
valid for a Gaussian distribution.
The $\la H_1^\perp\ra$ in Eq.~(\ref{apower-weighted}) can be taken from
Eq.~(\ref{apower}) (recall that the analyzing power (\ref{apower}) was
extracted under the assumption of a Gaussian transverse momentum distribution
\cite{Efremov:2001ia}).

Therefore we obtain the relation
\be\label{relation-diff-weights}
    A_{UT}^{\sin(\phi+\phi_S)k_\perp/\la P_{h\perp}\ra}
    = \beta_{\rm Gauss}\; A_{UT}^{\sin(\phi+\phi_S)} \;.\ee
The constant $\beta_{\rm Gauss}$ ``converts'' between the
differently weighted asymmetries and is given by
\be\label{beta_w}
    \beta_{\rm Gauss}
    = \frac{2}{\pi\,a_{\rm Gauss}}
    = \frac{4\la z\ra}{\pi} \;
    \sqrt{1+\la z^2\ra\la P_{{\rm N}\perp}^2\ra/\la P_{h\perp}^2\ra}
    \approx 0.55\; \ee
for the numbers in the HERMES experiment. Thus, in order to obtain our
prediction for $A_{UT}^{\sin(\phi+\phi_S)k_\perp/\la P_{h\perp}\ra}$ it is
sufficient to multiply the results in Figs.~2a and 2b by the factor $0.55$.
We stress that the ``conversion factor'' $\beta_{\rm Gauss}$
is model dependent. For a different model of transverse momenta
$\beta_{\rm model} \neq \beta_{\rm Gauss}$.
In particular, $\beta_{\rm model}$ could numerically be
different from the result in Eq.~(\ref{beta_w}).

Our prediction for $A_{UT}^{\sin(\phi+\phi_S)}$ is more robust
than that for $A_{UT}^{\sin(\phi+\phi_S)k_\perp/\la
P_{h\perp}\ra}(x)$ since the latter -- in addition to other
assumptions in our approach -- also tests the assumption of a
Gaussian transverse momentum distribution. In fact, the only
assumption entering our prediction for
$A_{UT}^{\sin(\phi+\phi_S)}$ -- and the analysis of
$A_{UL}^{\sin\phi}$ in Ref.~\cite{Efremov:2001ia} -- is that a
generic unintegrated fragmentation function $F(z,{\bf k}_T^2)$ can be
approximated by \be\label{factor-x-kT}
    F(z,{\bf k}_T^2) \approx F(z)\,G({\bf k}_T^2)\;,
\ee where $G({\bf k}_T^2)$ satisfies $\int\di^2{\bf k}_T\,G({\bf
k}_T^2) = 1$, and analogous for unintegrated distribution functions.
For a Gaussian distribution one sets 
$G({\bf k}_T^2)=\exp(-{\bf k}_T^2/\la {\bf k}_T^2\ra)/(\pi\la{\bf k}_T^2\ra)$. 
Assuming (\ref{factor-x-kT}) but taking a different
model for $G({\bf k}_T^2)$ we would obtain a {\sl  different}
constant $a_{\rm model}\neq a_{\rm Gauss}$ in Eq.~(\ref{a_Gauss}).
With a different model for transverse momenta, however, we also
would have obtained a different result in Eq.~(\ref{apower}) for
$\la H_1^\perp\ra_{\rm model}$. (In this context the $\la
H_1^\perp\ra$ in Eq.~(\ref{apower}) should be labelled $\la
H_1^\perp\ra_{\rm Gauss}$ for clarity.) Thus under the assumption
(\ref{factor-x-kT}) the relation
$A_{UT}^{\sin(\phi+\phi_S)}\propto a_{\rm model}\la
H_1^\perp\ra_{\rm model}$ is model independent. Therefore our
predictions for $A_{UT}^{\sin(\phi+\phi_S)}$ shown in Figs.~2a and
2b do not depend on the Gaussian model 
but rely solely on the the assumption
(\ref{factor-x-kT}). If the assumption (\ref{factor-x-kT}) held
one could discriminate between different models for the transverse
momentum distributions by considering different powers of
transverse momentum in the weight $\sin(\phi+\phi_S)|{\bf
k}_\perp|^n$ ($n=0,\,1$). Considering different weights could
provide interesting phenomenological insights. However, from a
strict theoretical point of view the weighting with an adequate power 
of $|{\bf k}_\perp|$ is preferable \cite{Boer:1997nt}.

\paragraph*{Preliminary SMC results.}
Though devoted to the HERMES experiment let us conclude this section with 
a comment on the preliminary SMC data reported in Ref.~\cite{Bravar:rq}.
In the SMC experiment indications were found that the transverse target 
spin asymmetry $\propto\sin\phi_c A_N$ with $A_N = 0.11\pm0.06$, 
where the Collins angle $\phi_c \equiv \phi+\phi_S-\pi$ 
(cf.\ \cite{Bravar:rq} for the precise definition of $A_N$).
Our approach yields $A_N= -0.12$, i.e.\ an asymmetry of 
opposite sign \cite{Efremov:2001cz,Efremov:2001ia}
(due to $\sin\phi_c = -\sin(\phi+\phi_S)$).
Considering the preliminary status of the data of Ref.~\cite{Bravar:rq}
it is not possible to draw any conclusions at this stage.

\section{COMPASS experiment}
\label{Sect:COMPASS-AUT-AUL}

\paragraph*{Transverse target spin asymmetry.}
The beam energy available at COMPASS is $E_{\rm beam}=160\,{\rm GeV}$
\cite{LeGoff:qn}. For the kinematic cuts we shall take
\be\label{exp-cuts-COMPASS}
        2\,{\rm GeV}^2 < Q^2 <  50\,{\rm GeV}^2  , \;\;\;
       15\,{\rm GeV}^2 < W^2 < 300\,{\rm GeV}^2  , \;\;\;
        0.05           < y   < 0.9                 \;,\;\;\;
                         x   < 0.4                 \;,\ee
and evaluate the distribution functions at $Q^2=10\,{\rm GeV}^2$.
We take $\la P_{h\perp}\ra\approx 0.4\;{\rm GeV}$ and $\la z\ra\approx 0.4$.
The latter means that we can use for $\la H_1^\perp\ra/\la D_1\ra$ the result
in Eq.~(\ref{apower}) -- if we assume that the ratio
$\la H_1^\perp\ra/\la D_1\ra$ is only weakly scale dependent in the
range of scales relevant in the HERMES and COMPASS experiments.
The estimate of $A_{UT}^{\sin(\phi+\phi_S)}$ obtained
in this way is shown in Fig.~3a.

In the HERMES experiment the analyzing power (assuming our approach) is
$H_1^\perp(z)/D_1(z) \approx a\,z$ where the constant $a\approx \frac13$
\cite{Efremov:2001cz}.
This means that $\la H_1^\perp\ra/\la D_1\ra \approx a \la z\ra$.
If such a pattern held also at COMPASS energies,
it would be preferable to choose a larger low-$z$ cut in order to increase
$\la z\ra$ and thus the analyzing power  $\la H_1^\perp\ra/\la D_1\ra$
(at the price of a lower statistics) \cite{LeGoff:qn}.
For a different $\la z\ra$ the results shown in Fig.~3a have to be
rescaled appropriately.

%
%
\begin{figure}[t!]\label{Fig:predictions-AUT-AUL-COMPASS}
\begin{tabular}{ccc}
    \hspace{-0.8cm}
    \includegraphics[width=5.5cm,height=5.5cm]{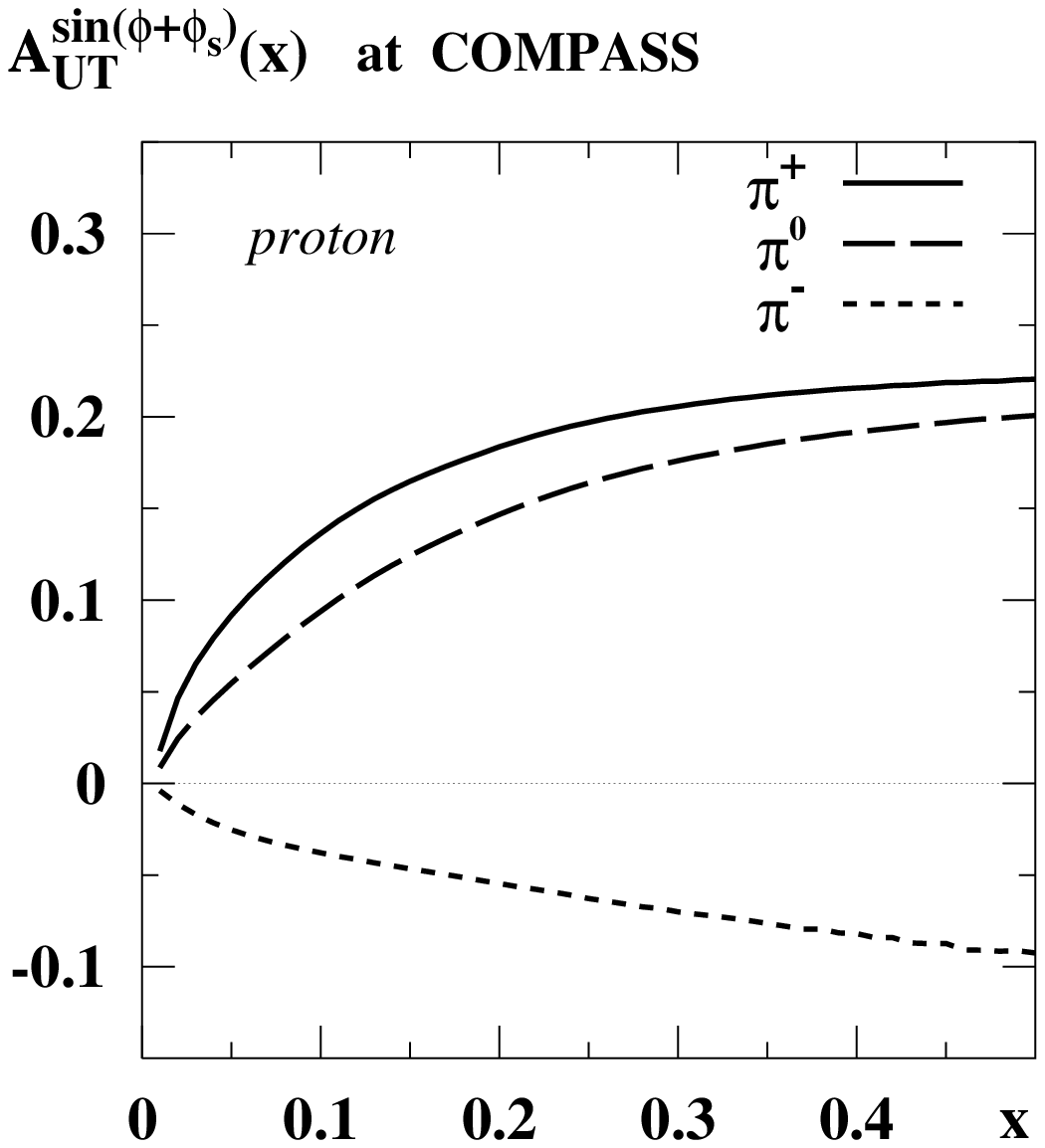}&
    \includegraphics[width=5.5cm,height=5.5cm]{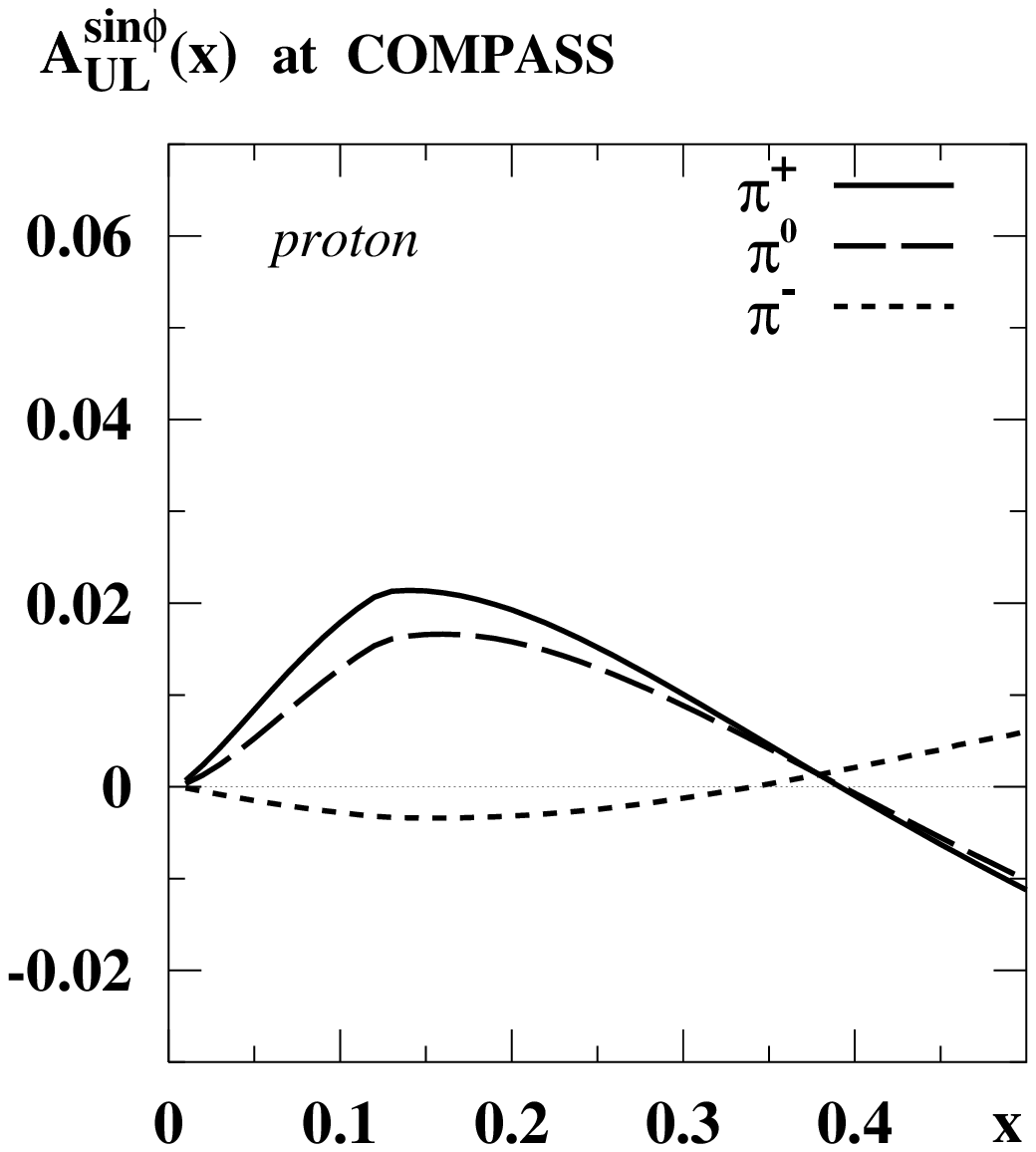}&
    \includegraphics[width=5.5cm,height=5.5cm]{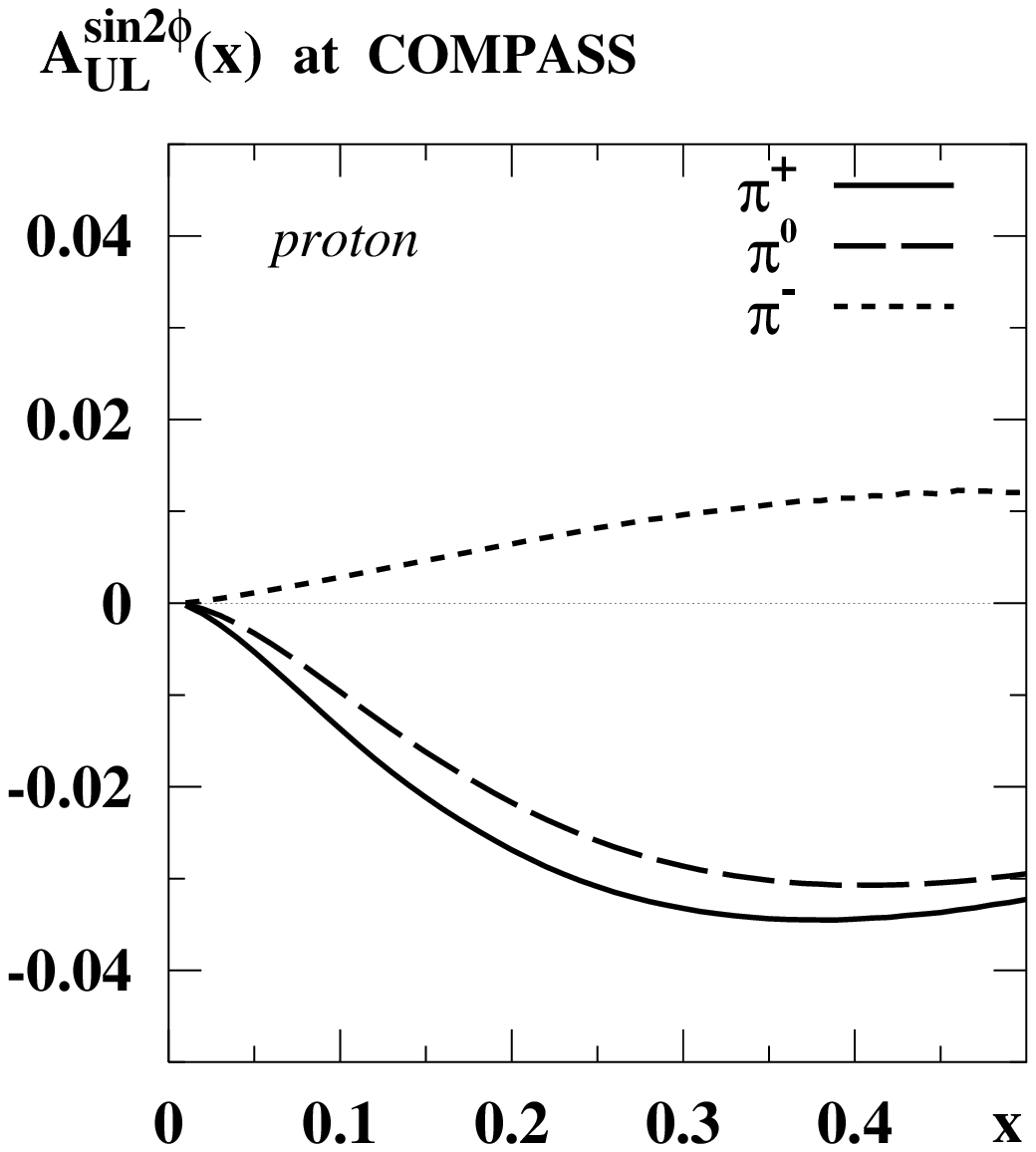}
    \cr
    \hspace{-1cm}
    \hspace{0.8cm} {\bf a} &
    \hspace{0.8cm} {\bf b} &
    \hspace{0.8cm} {\bf c}
\end{tabular}
    \caption{\footnotesize\sl
    {\bf a.}
    Prediction of the azimuthal asymmetry $A_{UT}^{\sin(\phi+\phi_S)}(x)$
    in SIDIS pion production from a transversely polarized proton target
    for the kinematics of the COMPASS experiment.
    Predictions of the azimuthal asymmetries $A_{UL}^{\sin\phi}(x)$
    ({\bf b}) and $A_{UL}^{\sin2\phi}(x)$ ({\bf c}) from a longitudinally
    polarized proton target for the kinematics of the COMPASS experiment.}
\end{figure}
%
%
%
%
\begin{figure}[t!]\label{Fig:predictions-AUT-AUL-COMPASS-deut}
\begin{tabular}{ccc}
    \hspace{-0.8cm}
    \includegraphics[width=5.5cm,height=5.5cm]{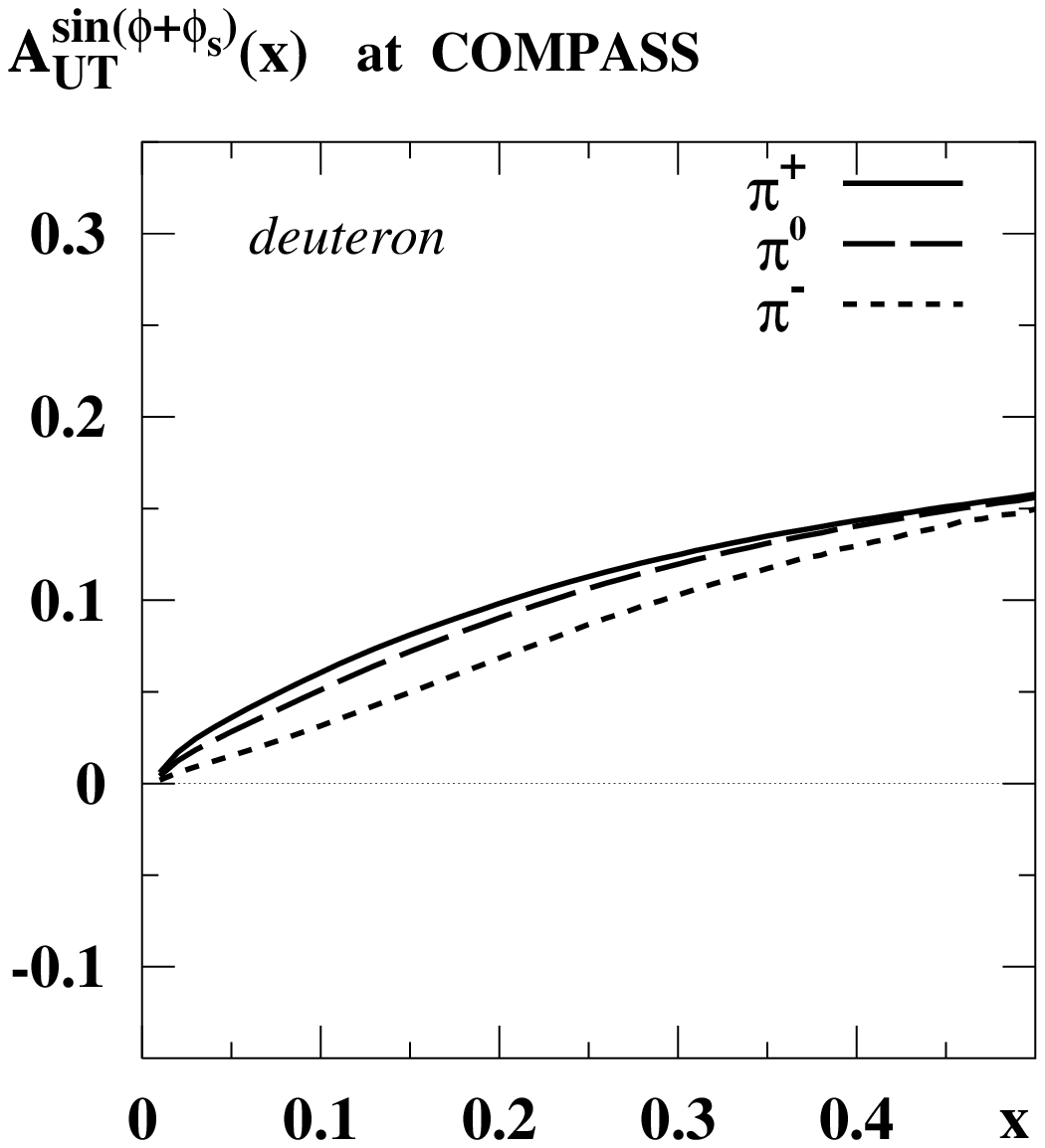}&
    \includegraphics[width=5.5cm,height=5.5cm]{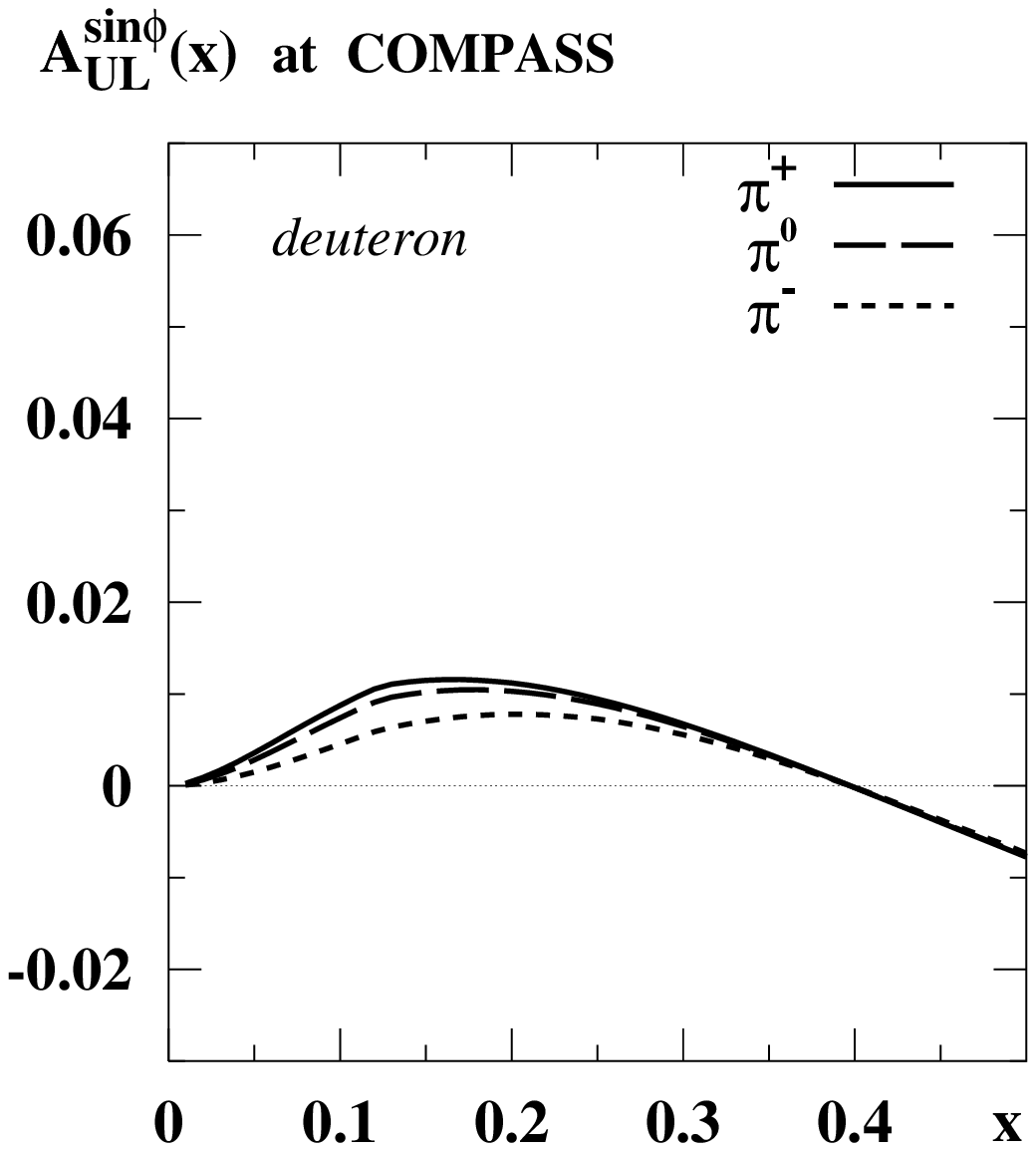}&
    \includegraphics[width=5.5cm,height=5.5cm]{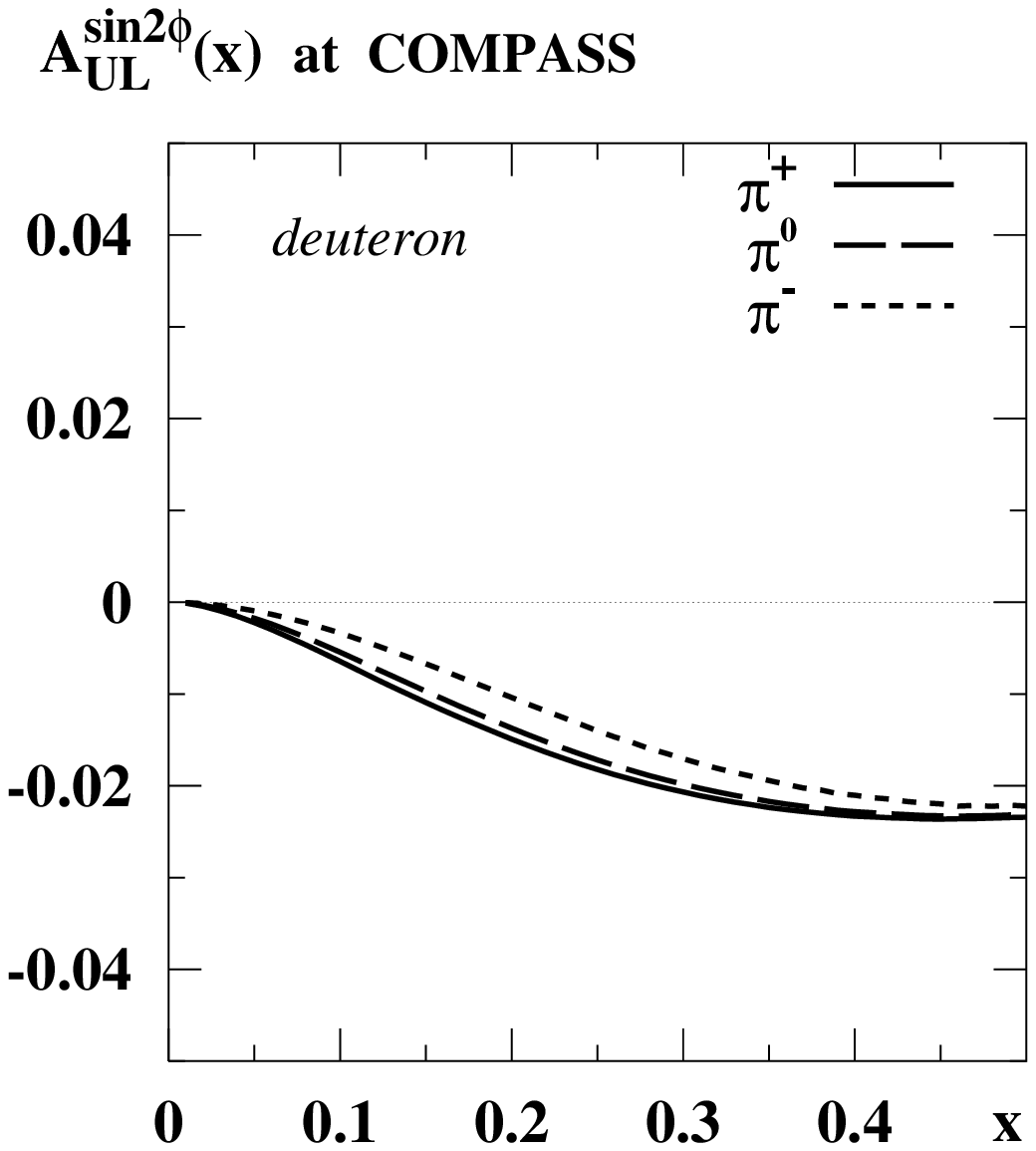}
    \cr
    \hspace{-1cm}
    \hspace{0.8cm} {\bf a} &
    \hspace{0.8cm} {\bf b} &
    \hspace{0.8cm} {\bf c}
\end{tabular}
    \caption{\footnotesize\sl
    The same as Fig.~3 but for the deuteron target.}
\end{figure}
%
%
Fig.~3a shows that $A_{UT}^{\sin(\phi+\phi_S)}$ can be of 
${\cal O}(20\%)$ at COMPASS energies, i.e.\  as large as at HERMES. 
This is not unexpected since this asymmetry is twist-2 (in the sense
that it is not power suppressed). Thus, the COMPASS experiment can
equally well shed some light on the dynamics of the Collins
effect. Actually, the theoretical accuracy of the predictions in
Fig.~3a is less than in the case of the predictions for HERMES
presented in the previous Section because one has to consider the
uncertainty introduced by assuming that the HERMES value for the
analyzing power (\ref{apower}) can be used at COMPASS energies.

\paragraph*{Longitudinal target spin asymmetries.}
About $80\%$ of the beam time the target polarization in the
COMPASS experiment will be longitudinal \cite{LeGoff:qn}. This
will allow to measure the longitudinal target spin asymmetries
$A_{UL}^{\sin\phi}$ and $A_{UL}^{\sin2\phi}$. (In the case of
longitudinal target polarization the azimuthal angle of the target
polarization vector $\phi_S=0$ or $\pi$, cf.\ Fig.~1.) The
estimates for these asymmetries in our approach are shown in
Figs.~3b and 3c. Clearly, the longitudinal target spin asymmetries
are much smaller than the transverse target spin asymmetry
$A_{UT}^{\sin(\phi+\phi_S)}$, however, 
the larger statistics could help to resolve them.

The $A_{UL}^{\sin2\phi}(x)$ asymmetry is of particular interest -- since 
it is one of the ``independent observables'' which could provide further
insights, cf.\ Section~\ref{Sect:Introduction}. This asymmetry was
found consistent with zero within error bars at HERMES
\cite{Avakian:rr,Airapetian:1999tv,Airapetian:2001eg,Airapetian:2002mf}.
In our approach at HERMES energies $A_{UL}^{\sin2\phi}={\cal O}(1\%)$
\cite{Efremov:2001cz,Efremov:2001ia} -- in agreement with the experiment. 
In the kinematics of the COMPASS experiment we find 
$A_{UL}^{\sin2\phi}(x)={\cal O}(3\%)$, i.e.\ of the order of 
magnitude of the $A_{UL}^{\sin\phi}(x)$ asymmetry observed at HERMES. 

\paragraph*{Deuteron target.}
The single target spin asymmetries  
$A_{UL}^{\sin(\phi+\phi_S)}$, $A_{UL}^{\sin2\phi}$ and $A_{UL}^{\sin\phi}$  
for the deuteron target at COMPASS energies are shown respectively in 
Figs.~4a, 4b and 4c. The deuteron asymmetries for $\pi^+$, $\pi^0$ and $\pi^-$
 are all of comparable order of magnitude and about half the magnitude of the 
$\pi^+$ proton asymmetries.

\section{Sivers effect azimuthal asymmetries}
\label{Sect:Sivers-effect}

Actually, our approach would imply the vanishing of
$A_{UT}^{\sin(\phi-\phi_S)}(x)$ asymmetry, which is due to the
Sivers effect \cite{Boer:1997nt} and will be measured at HERMES and COMPASS 
simultaneously with $A_{UT}^{\sin(\phi+\phi_S)}(x)$.
However, this cannot be taken literally as a
prediction for the following reason. The chiral quark-soliton model was 
derived from the instanton vacuum model and can be considered as the 
leading order in terms of the so-called instanton packing fraction 
$\frac {\rho}{R}\sim \frac{1}{3}$ ($\rho$ and $R$ are respectively 
the average size and separation of instantons in Euclidean space time).
In this order the T-odd distribution functions vanish. 
In higher orders the Sivers function can be well non-zero and all one can 
conclude at this stage is that the Sivers function is suppressed with respect
to the T-even\footnote{
    The suppression of T-odd with respect to T-even distributions is
    natural. E.g.\ in the quark-diquark models with gluon exchange
    \cite{Brodsky:2002cx,Brodsky:2002pr,Gamberg:2003ey} -- where the
    Sivers function was ``rediscovered'', cf.\  \cite{Collins:2002kn} --
    T-even distributions appear at the tree-level while T-odd ones
    appear only at one-loop level. Thus, whatever (small) parameter
    justifies the perturbative calculation of distribution functions in
    the quark-diquark model, it generically suppresses T-odd distributions
    with respect to T-even ones.}
twist-2 distribution functions $f_1^a(x)$, $g_1^a(x)$ and
$h_1^a(x)$.\footnote{In the case of the pure twist-3
$\widetilde{h}_L^a(x)$ \cite{Dressler:1999hc} (or
$\widetilde{g}_T^a(x)$ \cite{Balla:1997hf}) it was shown on the
basis of ~\cite{Diakonov:1996qy} that the suppression in the
instanton medium with respect to twist-2 distributions is very
strong.} However, considering that $H_1^\perp(z)$ is much smaller
than $D_1(z)$, cf.\ Eq.~(\ref{apower}), it is questionable whether
such a suppression could be sufficient such that in physical cross
sections the Collins effect $\propto h_1^a(x)H_1^\perp(z)$ is
dominant over the Sivers  effect $\propto f_{1T}^\perp(x)D_1(z)$.
In Ref.~\cite{Efremov:2003tf} it was estimated that for the
particular case of $A_{UL}^{\sin\phi}$ asymmetries in the HERMES
kinematics this still could be true: Using the Sivers function of
Ref.~\cite{Anselmino:1994tv} fitted to explain the E704 data
\cite{Adams:1991cs} on single spin asymmetries in
$pp^\uparrow\to\pi X$ solely in terms of the Sivers effect, it was
shown that the Sivers effect could give rise to
$A_{UT}^{\sin(\phi-\phi_S)}={\cal O}(10\%)$ while its contribution
to $A_{UL}^{\sin\phi}$ is negligible with respect to the Collins
effect. Of course, the E704 data need not to be due to the Sivers
effect alone, and the Sivers effect in $pp^\uparrow\to\pi X$ need
not to be simply related to the Sivers effect in SIDIS.\footnote{
    Cf.\ the corresponding discussions of the Sivers effect in SIDIS
    and the Drell-Yan process, where the Sivers functions differ by
    an overall sign \cite{Collins:2002kn,Belitsky:2002sm}.}
Therefore the observation of Ref.~\cite{Efremov:2003tf} has to be
considered with care. Interestingly, in the quark-diquark model
one finds a comparably large $A_{UT}^{\sin(\phi-\phi_S)} = {\cal O}(10\%)$
\cite{Gamberg:2003ey}.

To summarize, though in our approach the Sivers functions vanishes,
there need not be a contradiction if Sivers effect asymmetry
$A_{UT}^{\sin(\phi-\phi_S)}$ would be observed to be sizable.
The measurements of $A_{UT}^{\sin(\phi\pm\phi_S)}$
at HERMES and COMPASS (and $A_{UL}^{\sin2\phi}$ at COMPASS)
will clarify the situation.

\section{Conclusions}
\label{Sect:Conclusions}

Recently HERMES observed noticeable azimuthal single spin asymmetries 
$A_{UL}^{\sin\phi}$ in SIDIS off a longitudinally polarized target 
\cite{Avakian:rr,Airapetian:1999tv,Airapetian:2001eg,Airapetian:2002mf}.
These asymmetries could arise from both the Collins and the Sivers effect
and are therefore difficult to interpret. 
Important further insights can be gained from the study of azimuthal
asymmetries in SIDIS off a transversely polarized target because the 
angular distribution of the produced pions allows to cleanly distinguish 
between the Collins and Sivers effect \cite{Mulders:1995dh,Boer:1997nt}.

In this note we have presented estimates of 
the azimuthal single spin asymmetries due to the Collins effect,
$A_{UT}^{\sin(\phi+\phi_S)}$, both for the HERMES and COMPASS experiments. 
These calculations are based on two ingredients. One ingredient,
which is responsible for the $x$-shape of the predicted asymmetries, 
is the chirally odd transversity distribution function $h_1^a(x)$ 
provided by the chiral quark-soliton model ($\chi$QSM) \cite{h1-model}. 
The sign and the overall normalization of the predicted $A_{UT}$ asymmetries 
are fixed by the second ingredient, namely by properties of the Collins
fragmentation function $H_1^\perp$ resulting from our analysis
\cite{Efremov:2001cz} of the $A_{UL}^{\sin\phi}$ asymmetries observed 
in the HERMES experiment. On the basis of this approach we estimate the 
$A_{UT}^{\sin(\phi+\phi_S)}$ to be about $20\%$ for $\pi^+$ and $\pi^0$ 
from a proton target and roughly $10\%$ for all pions from a deuterium target.

Choosing another weight, namely $\sin(\phi-\phi_S)$, it is
possible to project out another azimuthal asymmetry which is due
to the Sivers effect only \cite{Boer:1997nt}. If taken {\sl literally}, 
our approach would predict a vanishing Sivers effect asymmetry 
$A_{UT}^{\sin(\phi-\phi_S)}$ because in the $\chi$QSM the Sivers distribution 
function vanishes. This shortcoming is met basically in all chiral effective 
models \cite{Pobylitsa:2002fr} and reflects the limitations of such models to 
describe T-odd distribution functions. In the $\chi$QSM, 
which is based on an expansion in terms of the packing
fraction of the instantons in the vacuum, T-odd distribution functions
are subleading quantities in contrast T-even distribution functions.
However, a Sivers function as large as obtained in the quark-diquark models 
with gluon exchange \cite{Brodsky:2002cx,Brodsky:2002pr,Gamberg:2003ey}
yielding $A_{UT}^{\sin(\phi-\phi_S)}={\cal O}(10\%)$ \cite{Gamberg:2003ey} 
would not be in contradiction with our approach \cite{Efremov:2003tf}.

Noteworthy, the longitudinal target polarization program of the
COMPASS experiment may also well contribute to the understanding
of single spin asymmetries in SIDIS. Our approach predicts the
$A_{UL}^{\sin2\phi}$ asymmetry, which was found consistent with
zero within (relatively large) error bars at HERMES, is of 
${\cal O}(3\%)$ in the COMPASS kinematics and can probably be resolved.
This asymmetry is due to the Collins effect only and its
measurement would provide valuable independent information. The
$A_{UL}^{\sin\phi}$ asymmetry is about $(1-2)\%$ and more difficult
to measure for COMPASS.

A measurement of the $A_{UT}^{\sin(\phi+\phi_S)}$ asymmetry at HERMES and 
COMPASS and the $A_{UL}^{\sin2\phi}$ at COMPASS of comparable magnitude as we 
estimated here would support the observation \cite{Efremov:2003tf} that the 
Sivers effect could play a sub-dominant role in the $A_{UL}^{\sin\phi}$ 
asymmetries measured by HERMES
\cite{Avakian:rr,Airapetian:1999tv,Airapetian:2001eg,Airapetian:2002mf} 
and a posteriori justify the attempts  \cite{DeSanctis:2000fh,Anselmino:2000mb,Efremov:2000za,Ma:2002ns,Korotkov:1999jx,Efremov:2001cz,Efremov:2001ia}
to interpret these data in terms of the Collins effect only.
In contrast, deviations from our predictions could provide valuable hints 
how those attempts should be modified.  We will -- in any case -- soon learn 
a lot from the HERMES and COMPASS experiments.

\vspace{0.5cm}
\paragraph*{Acknowledgement.}
We are grateful to H.~Avakian, M.~Beckmann, I.~Ludwig, R.~Seidl 
for fruitful discussions.
A.~E.\ is partially supported by INTAS grant 00/587
and RFBR grant 03-02-16816 and DFG-RFBR 03-02-04022.
This work has partly been performed under the contract
HPRN-CT-2000-00130 of the European Commission. The work is partially
supported by BMBF and DFG of Germany and by the COSY-Juelich project.

\appendix
\section{Expressions for longitudinal target polarization asymmetries}

For the convenience of the reader we summarize the expressions
for $A_{UL}^{\sin2\phi}$ and $A_{UL}^{\sin\phi}$ which were derived
in \cite{Efremov:2000za,Efremov:2001cz,Efremov:2001ia} on the basis 
of the results from Ref.~\cite{Mulders:1995dh}:
\ba
\label{AUL-sinPhi}
    A_{UL}^{\sin\phi}(x) &=& a_{\rm Gauss} \, \Biggl(
    P_L(x)\;\frac{\sum_a e_a^2\,x^2 h_L^a(x)\,\la H_1^{\perp a}\ra}
                     {\sum_b e_b^2\,x   f_1^b(x)\,\la D_1^b\ra\,}
      + P_1(x)\;\frac{\sum_a e_a^2\,x   h_1^a(x)\,\la H_1^{\perp a}\ra}
                 {\sum_b e_b^2\,x   f_1^b(x)\,\la D_1^b\ra\,}\Biggr)
    \\
\label{AUL-sin2Phi}
        A_{UL}^{\sin 2\phi}(x) &=&
    4\la z\ra^2 a_{\rm Gauss}^2 \frac{2M_N}{\la P_{\perp h}\ra}\; P_2(x)\;
    \frac{\sum_a e_a^2\,2x^3 \int_x^1\di y h_1^a(y)/y^2\,\la H_1^{\perp a}\ra}
         {\sum_b e_b^2\, x   f_1^b(x)\,\la D_1^b\ra}
\ea
where $a_{\rm Gauss}$ is defined as in Eq.~(\ref{a_Gauss}) and the $P_i$
($i=L,\;1,\;2$) are given by
\ba\label{A-prefactors-def}
   P_L(x)&=&\frac{2\int\!\di y\,2(2-y)\,\sqrt{1-y}\;\cos\theta_\gamma{M_N}/Q^5}
                 {\int\!\di y\,(1-y+y^2/2)\,/\,Q^4}\;\;,\nonumber\\
   P_1(x)&=&-\;\frac{2\int\!\di y\,(1-y)\,\sin\theta_\gamma/Q^4}
                    {\int\!\di y\,(1-y+y^2/2)\,/Q^4}\;\;,\nonumber\\
   P_2(x)&=& \frac{2\int\!\di y\,(1-y)\,\cos\theta_\gamma/Q^4}
                  {\int\!\di y\,(1-y+y^2/2)\,/Q^4}\;.\ea


\end{document}